\pdfoutput=1

\documentclass[runningheads]{llncs}

\usepackage{cite}
\usepackage{graphicx}
\usepackage[breaklinks]{hyperref}
\usepackage{xurl}
\usepackage{acronym}
\usepackage{balance}
\usepackage{booktabs}
\usepackage{tabularx}
\usepackage[table,xcdraw]{xcolor}
\usepackage{rotating}
\usepackage{multirow}
\usepackage{makecell}
\usepackage{pifont}
\usepackage{fixltx2e}
\usepackage{microtype}
\usepackage{xcolor}
\usepackage{enumitem}
\usepackage{wasysym}
\usepackage{tikz}

\newcommand\copyrighttext{%
	\footnotesize The final publications is available at Springer via \url{https://doi.org/10.1007/978-3-030-44081-7_8}}
	\newcommand\copyrightnotice{%
	\begin{tikzpicture}[remember picture,overlay]
	\node[anchor=south,yshift=10pt] at (current page.south) {\fbox{\parbox{\dimexpr\textwidth-\fboxsep-\fboxrule\relax}{\copyrighttext}}};
	\end{tikzpicture}%
}

\newcolumntype{A}[1]{>{\raggedright\arraybackslash}p{#1}}
\newcolumntype{B}[1]{>{\centering\arraybackslash}p{#1}}

\acrodef{cdn}[CDN]{Content Delivery Network}
\acrodef{cp}[CP]{Content Provider}
\acrodef{eool}[EOOL]{End of option list}
\acrodef{noop}[NOOP]{No-Operation}
\acrodef{mss}[MSS]{Maximum Segment Size}
\acrodef{isn}[ISN]{Initial Sequence Number}
\acrodef{bsd}[BSD]{Berkeley Software Distribution}
\acrodef{os}[OS]{Operating System}
\acrodefplural{oss}[OSs]{Operating Systems}
\acrodef{ecn}[ECN]{Explicit Congestion Notification}
\acrodef{dscp}[DSCP]{differentiated services code point}
\acrodef{mtu}[MTU]{maximum transmission unit}
\acrodef{alp}[ALP]{application layer protocol}
\acrodef{grease}[GREASE]{Generate Random Extensions And Sustain Extensibility}
\acrodef{sack}[SACK]{Selective Acknowledgment}

\definecolor{colorGreenCheckmark}{rgb}{0.25, 0.56, 0.35}
\definecolor{colorRedXmark}{rgb}{0.76, 0.23, 0.13}

\newcommand{\eg}{e.g., }

\newcommand{\ie}{i.e., }

\newcommand{\cmark}{\textcolor{colorGreenCheckmark}{\ding{51}}}
\newcommand{\xmark}{\textcolor{colorRedXmark}{\ding{55}}}
\newcolumntype{Y}{>{\centering\arraybackslash}X}
\newcommand{\ftarget}{F\textsubscript{Target}}
\newcommand{\fpath}{F\textsubscript{Path}}
\newcommand{\ca}{\raise.17ex\hbox{$\scriptstyle\sim$}}

\newcommand{\afblock}[1]{\noindent{\textbf{#1 }}}
\newcommand{\takeaway}[1]{\noindent{\textit{\textbf{Takeaway:}}} \textit{#1}}

\setitemize{labelsep=2.5pt,leftmargin=4mm}

\newcommand{\checksumIncorrect}{\textit{ChecksumIncorrect }}
\newcommand{\checksumZero}{\textit{ChecksumZero }}
\newcommand{\optionSupport}{\textit{OptionSupport }}
\newcommand{\optionUnknown}{\textit{OptionUnknown }}
\newcommand{\mssSupport}{\textit{MSSSupport }}
\newcommand{\mssMissing}{\textit{MSSMissing }}
\newcommand{\reserved}{\textit{Reserved }}
\newcommand{\urgentPointer}{\textit{UrgentPointer }}

\begin{document}
\title{MUST, SHOULD, DON'T CARE:\\TCP Conformance in the Wild}
\titlerunning{MUST, SHOULD, DON'T CARE: TCP Conformance in the Wild}  

\authorrunning{M. Kosek et al.}
\tocauthor{Mike Kosek, Leo Bl\"ocher, Jan R\"uth, Torsten Zimmermann, Oliver Hohlfeld}

\author{Mike Kosek \and Leo Bl\"ocher \and Jan R\"uth \and Torsten Zimmermann \and \newline Oliver Hohlfeld$^\dagger$}
\institute{Communication and Distributed Systems, RWTH Aachen University\\
\email{\{kosek, bloecher, rueth, zimmermann\}@comsys.rwth-aachen.de}\\
$^\dagger$Chair of Computer Networks, Brandenburg University of Technology\\
\email{oliver.hohlfeld@b-tu.de}}

\maketitle              \begin{abstract}
Standards govern the SHOULD and MUST requirements for protocol implementers for interoperability.
In case of TCP that carries the bulk of the Internets' traffic, these requirements are defined in RFCs.
While it is known that not all optional features are implemented and non-conformance exists, one would assume that TCP implementations at least conform to the minimum set of MUST requirements.
In this paper, we use Internet-wide scans to show how Internet hosts and paths conform to these basic requirements.
We uncover a non-negligible set of hosts and paths that do not adhere to even basic requirements.
For example, we observe hosts that do not correctly handle checksums and cases of middlebox interference for TCP options.
We identify hosts that drop packets when the urgent pointer is set or simply crash.
Our publicly available results highlight that conformance to even fundamental protocol requirements should not be taken for granted but instead checked regularly.
\copyrightnotice

\end{abstract}
\section{Introduction}
\label{sec:intro}

Reliable, interoperable, and secure Internet communication largely depends on the adherence to standards defined in RFCs.
These RFCs are simple text documents, and any specifications published within them are inherently informal, flexible, and up for interpretation, despite the usage of keywords indicating the requirement levels~\cite{bradner:rfc2119:Keywords}, \eg SHOULD or MUST.
It is therefore expected and known that violations---and thus non-conformance---do arise unwillingly.
Nevertheless, it can be assumed that Internet hosts widely respect at least a minimal set of mandatory requirements.
To which degree this is the case is, however, unknown.

In this paper, we shed light on this question by performing Internet-wide active scans to probe if Internet hosts and paths are conformant to a set of \emph{minimum} TCP requirements that any TCP speaker MUST implement.
This adherence to the fundamental protocol principles is especially important since TCP carries the bulk of the data transmission in the Internet.
The basic requirements of a TCP host are defined in RFC\,793~\cite{unknown:rfc793:TransmissionControlProtocol}---the core TCP specification.
Since its over 40 years of existence, it has accumulated over 25 accepted errata described in RFC\,793bis-Draft14~\cite{eddy:rfc793Draft14:tcpTransmissionControlProtocolSpecification}, which is a draft of a planned future update of the TCP specification, incorporating all minor changes and errata to RFC\,793. 
We base our selection of probed requirements on formalized MUST requirements defined in this drafted update to RFC\,793.

The relevance of TCP in the Internet is reflected in the number of studies assessing its properties and conformance. 
Well studied are the interoperability of TCP extensions~\cite{cardwell:usenix13:packetdrill}, or within special purpose scenarios~\cite{marinos:sigcomm17:DiskCryptNet,marinos:sigcomm14:NetworkStackPerformanceSpecialization}, and especially non-conformance introduced by middleboxes on the path~\cite{detal:imc13:tracebox,kuehlewind:tma18:tracingInternetPathTransparency}.
However, the conformance to basic mandatory TCP features has not been studied in the wild.
We close this gap by studying to which degree TCP implementations in the wild conform to MUST requirements.
Non-conformance to these requirements limits interoperability, extensibility, performance, or security properties, leading to the essential necessity to understand who does not adhere to which level of non-conformance.
Uncovering occurrences of non-conformities hence reveal areas of improvement for future standards.
A recent example is QUIC, where effort is put into the avoidance of such misconceptions during standardization \cite{piraux:epiq18:QuicImplementationObserving}.

With our large scale measurement campaign presented in this paper, we show that while the majority of end-to-end connections are indeed conforming to the tested requirements, a non-trivial number of end-hosts as well as end-to-end paths show non-conformities, breaking current and future TCP extensions, and even voiding interoperability thus reducing connectivity. We show that
\begin{itemize}
	\item In a controlled lab study, non-conformance already exists at the OS-level: only two tested stacks (Linux and lwIP) pass all tests, where, surprisingly, others (including macOS and Windows) fail in at least one category each.
Observing non-conformance in the wild can therefore be expected.
	\item In the wild, we indeed found a non-negligible amount of non-conformant hosts.
	For example, checksums are not verified in \ca3.5\% of cases, and middleboxes inject non-conformant MSS values.
	Worrisome, using reserved flags or setting the urgent pointer can render the target host unreachable.
\item At a infrastructure level, 4.8\% of the Alexa domains with and without \texttt{www.}\ prefix show different conformance levels (e.g., because of different infrastructures: CDN vs.\ origin server), mostly due to flags that limit reachability.
	The reachability of websites can thus depend on the \texttt{www.}\ prefix.
\end{itemize}

\afblock{Structure.} In Section \ref{sec:rw} we present related work followed by our methodology and its validation in Section~\ref{sec:methodogolgy}.
The design and evaluation of our Internet-wide TCP conformance scans are discussed in Section~\ref{sec:results} before we conclude the paper.

\section{Related Work}
\label{sec:rw}
Multiple measurement approaches have focused on the conformance of TCP implementations on servers, the presence of middleboxes and their interference on TCP connections, and non-standard conform behavior.
In the following, we discuss similarities and differences of selected approaches to our work.

\afblock{TCP Stack Behavior.}
One line of research aims at characterizing remote TCP stacks by their behavior (e.g., realized in the TCP Behavior Inference Tool (\texttt{TBIT})~\cite{pahdye:sigcomm01:tbit} in 2001).
One aspect is to study the deployment of TCP tunings (e.g., the initial window configuration~\cite{iwpaper, iwTNSM, rueth:tma18:DemystifyCdnIw}) or TCP extensions (e.g., Fast Retransmit~\cite{pahdye:sigcomm01:tbit}, Fast Open~\cite{tfoInitialDeployment, tfoChristoph}, \ac{sack}~\cite{pahdye:sigcomm01:tbit, MedinaCCR2005, MirjaECN}, or \ac{ecn}~\cite{pahdye:sigcomm01:tbit, MedinaIMC2004, MedinaCCR2005, LangleyECN, ECNReadiness, MirjaECN} and ECN++\cite{BobECNPlus} to name a few).
While these works aim to generally characterize stacks by behavior and to study the availability and deployability of TCP extensions, our work specifically focuses on the \emph{conformance} of \emph{current} TCP stacks to \emph{mandatory} behavior every stack must implement.
A second aspect concerns the usage of behavioral characterizations to \emph{fingerprint} TCP stacks (e..g., via active~\cite{nmapFingerprinting} or passive~\cite{RobPassiveTCPFingerprinting} measurements) and mechanisms to defeat fingerprinting (e.g.,~\cite{defeatingFingerprinting}).

\afblock{Middlebox Interference.}
The end-to-end behavior of TCP not only depends on the stack implementations, but also on on-path middleboxes~\cite{rfc3234}, which can tune TCP performance and security but also (negatively) impact protocol mechanisms and extensions (see e.g.,~\cite{MedinaIMC2004, MedinaCCR2005, ECNReadiness}).
Given their relevance, a large body of work studies the impact within the last two decades and opens the question if TCP is still extensible in today's Internet.
Answering this question resulted in a methodology for middlebox inference which is extended by multiple works to provide middlebox detection tools to assess their influence;
By observing the differences between sent and received TCP options at controlled endpoints (\texttt{TCPExposure}~\cite{honda:imc11:ExtendTcp}), it is observed that 25\,\% of the studied paths tamper with TCP options, \eg with TCP's \ac{sack} mechanism.
Similarly, \texttt{tracebox}~\cite{detal:imc13:tracebox} also identifies middleboxes based on modifications of TCP options, but as client-side only approach without requiring server control.
Besides also identifying the issues with TCP's \ac{sack} option, they highlight the interference with TCP's MSS option and the incorrect behavior of TCP implementations when probing for MPTCP support.
\texttt{PATHSpider}~\cite{kuehlewind:tma18:tracingInternetPathTransparency} extends \texttt{tracebox} to test more TCP options, \eg \ac{ecn} or \ac{dscp}.
They evaluate their tool in an \ac{ecn} support study, highlighting that some intermediaries tamper with the respective options, making a global \ac{ecn} deployment a challenging task.
Further investigating how middleboxes harm TCP traffic, a tracebox-based study~\cite{edeline:tma19:TransportOssification} shows that more than a third of all studied paths cross at least one middlebox, and that on over 6\% of these paths TCP traffic is harmed.
Given the negative influence of transparent middleboxes, proposals enable endpoints to identify and negotiate with middleboxes using a new TCP option~\cite{middleboxNegotiation} and to generally cooperate with middleboxes~\cite{TCPHiccups}.
While we focus on assessing TCP conformance to mandatory behavior, we follow tracebox's approach to differentiate non-conforming stacks from middlebox interference causing non-conformity.

\takeaway{While a large body of work already investigates TCP behavior and middlebox inference, a focus on conformance to mandatory functionality required to implement is missing---a gap that we address in this study.}

\section{Methodology}
\label{sec:methodogolgy}

We test TCP conformance by performing active measurements that probe for mandatory TCP features and check adherence to the RFC.
We begin by explaining how we detect middleboxes before we define the test cases and then validate our methodology in controlled testbed experiments.

\subsection{Middlebox Detection}
Middleboxes can alter TCP header information and thereby cause non-conformance, which we would wrongly attribute to the probed host without performing a middlebox detection.
Therefore, we use the tracebox approach~\cite{detal:imc13:tracebox} to detect interfering middleboxes by sending and repeating our probes with increasing IP TTLs.
That is, in every test case (see Section~\ref{sec:design}), the first segment is sent multiple times with increasing TTL values from 1 to 30 in parallel while capturing ICMP time exceeded messages.
We limit the TTL to 30 since we did not observe higher hop counts in our prior work for Internet-wide scans~\cite{rueth:pam19:hiddentreasures}.
To distinguish the replied messages and determine the hop count, we encode the TTL in the IPv4 ID and in the TCP acknowledgment number, window size, and urgent pointer fields.
We chose to encode the TTL in multiple header fields since middleboxes could alter every single one. 
These repetitions enable us to pinpoint and detect (non-)conformance within the end-to-end path if ICMP messages are issued by the intermediaries quoting the expired segment.
Please note that alteration or removal of some of our encodings does \emph{not} render the path or the specific hop non-conformant.
A non-conformance is only attested, if the actual tested behavior was modified as visible through the expired segment.
Further, since only parts of the fields---all 16 or 32 bits in size---may be altered by middleboxes (\eg slight changes to the window size), we repeat each value as often as possible within every field.
Our TTL value of at most 30 can be encoded in 5 bits, and thus be repeated 3 to 6 times in the selected fields.
Additionally, the TCP header option \ac{noop} allows an opaque encoding of the TTL.
Specifically, we append as many \acp{noop} as there are hops in the TTL to the fixed-size header.
Other header fields are either utilized for routing decisions (\eg port numbers in load balancers) or are not opaque (\eg sequence numbers), rendering them unsuitable.
Depending on the specific test case, some of the fields are not used for the TTL encoding.
For example, when testing for urgent pointer adherence, we do not encode the TTL in the urgent pointer field.

\subsection{TCP Conformance Test Cases}
\label{sec:design}
Our test cases check for observable TCP conformance of end-to-end connections by actively probing for a set of \emph{minimum} requirements that any TCP must implement. 
We base our selection on 119 explicitly numbered requirements specified in RFC\,793bis-Draft14~\cite{eddy:rfc793Draft14:tcpTransmissionControlProtocolSpecification}, of which 69 are absolute requirements (\ie \emph{MUSTs}\cite{bradner:rfc2119:Keywords}). 
These MUSTs resemble minimum requirements for \emph{any} TCP connection participating in the Internet---not only for hosts, but also for intermediate elements within the traversed path. 
The majority of these 69 MUSTs address internal state-handling details, and can therefore not be observed or verified via active probing.
To enable an Internet-wide assessment of TCP conformance, we thus focus on MUST requirements whose adherence is \emph{observable} by communicating with the remote host.
We synthesize eight tests from these requirements, which we summarize in Table~\ref{tab:requirements}, and discuss them in the following paragraphs.
Each test is in some way critical to interoperability, security, performance, or extensibility of TCP.
The complexity involved in verifying conformance to other advanced requirements often leads to the exclusion of these seemingly fundamental properties in favor of more specialized research.

\begin{table}[t]
	\small
	\setlength{\aboverulesep}{0pt}
   	\setlength{\belowrulesep}{0pt}
    	\setlength\extrarowheight{0.5pt}
\centering
	\rowcolors{2}{gray!15}{white}
	\begin{tabularx}{\textwidth}{B{11em}|X}
	\rowcolor{white!100}
	
	\multicolumn{1}{c|}{\textbf{Checksum}}	&\multicolumn{1}{c}{\textbf{PASS Condition}}\\ \Xhline{0.75pt}
	\multirow{3}{*}{\checksumIncorrect	(2,3)} & \begin{minipage}[t]{\linewidth}\begin{itemize}[nosep,after=\strut]\item When sending a SYN or an ACK segment with a non-zero but invalid checksum, a target must respond with a RST segment or ignore it\end{itemize}\end{minipage}\\\hline 
 	\checksumZero (2,3) 				 & \begin{minipage}[t]{\linewidth}\begin{itemize}[nosep,after=\strut]\item As above but with an explicit zeroed checksum\end{itemize}\end{minipage}\\\Xhline{0.75pt}
	\end{tabularx}
	\vspace{7.5pt}
	
	\begin{tabularx}{\textwidth}{B{11em}|X}
	\rowcolor{white!100}
	\multicolumn{1}{c|}{\textbf{Options}}	&\multicolumn{1}{c}{\textbf{PASS Condition}}\\ \Xhline{0.75pt}
	\multirow{2}{*}{\optionSupport (4)}  		& \begin{minipage}[t]{\linewidth}\begin{itemize}[nosep,after=\strut]\item When sending a SYN segment with \ac{eool} and \ac{noop} options, a target must respond with a SYN/ACK segment\end{itemize}\end{minipage}\\\hline 
	\multirow{2}{*}{\optionUnknown (6)} 		& \begin{minipage}[t]{\linewidth}\begin{itemize}[nosep,after=\strut]\item When sending a SYN segment with an unassigned option (\#\,158), a target must respond with a SYN/ACK segment\end{itemize}\end{minipage}\\\hline 
	\multirow{2}{*}{\mssSupport (4,14,16)}	& \begin{minipage}[t]{\linewidth}\begin{itemize}[nosep,after=\strut]\item When sending a SYN segment with an \ac{mss} of 515 byte, a target must not send segments exceeding 515 byte\end{itemize}\end{minipage}\\\hline 
	\multirow{3}{*}{\mssMissing (15,16)}		& \begin{minipage}[t]{\linewidth}\begin{itemize}[nosep,after=\strut]\item When sending a SYN segment without an \ac{mss}, a target must not send segments exceeding 536 byte (IPv4) or 1220 byte (IPv6, not tested)\end{itemize}\end{minipage}\\\Xhline{0.75pt}
	\end{tabularx}
	\vspace{7.5pt}
	
	\begin{tabularx}{\textwidth}{B{11em}|X}
	\rowcolor{white!100}
	\multicolumn{1}{c|}{\textbf{Flags}}	&\multicolumn{1}{c}{\textbf{PASS Condition}}\\ \Xhline{0.75pt}
	\multirow{6}{*}{\reserved (no MUST)}	 & \begin{minipage}[t]{\linewidth}\begin{itemize}[nosep,after=\strut]\item When Sending a SYN segment with a reserved flag set (\#\,2), a target must respond with a SYN/ACK segment with zeroed reserved flags
								    \item Subsequently, when sending an ACK segment with a reserved flag set (\#\,2), a target must not retransmit the SYN/ACK segment\end{itemize}\end{minipage}\\\hline 
	\multirow{2}{*}{\urgentPointer (30,31)} & \begin{minipage}[t]{\linewidth}\begin{itemize}[nosep,after=\strut]\item When sending a sequence of segments flagged as urgent, a target must acknowledge them with an ACK segment\end{itemize}\end{minipage}\\\Xhline{0.75pt}
	\end{tabularx}
	
	\vspace{2.5pt}
	\caption{Requirements based on the MUSTs (number from RFC shown in brackets) as defined in RFC\,793bis, Draft 14~\cite{eddy:rfc793Draft14:tcpTransmissionControlProtocolSpecification}.
	Further, we show the precise test sequence and the condition leading to a PASS for the test.}
	\label{tab:requirements}
	\vspace{-2em}
\end{table}

\afblock{TCP Checksum.}
The TCP checksum protects against segment corruption in transit and is mandatory to both calculate and verify.
Even though most Layer\,2 protocols already protect against segment corruption, it has been shown~\cite{stone:sigcomm00:checksums} that software or hardware bugs in intermediate systems may still alter packet data, and thus, high layer checksums are still vital.
Checksums are an essential requirement to consider due to the performance implications of having to iterate over the entire segment after receiving it, resulting in an incentive to skip this step even though today this task is typically offloaded to the NIC.
Both the \checksumIncorrect and the \checksumZero test (see Table~\ref{tab:requirements}) verify the handling of checksums in the TCP header.
They differ only in the kind of checksum used; the former employs a randomly chosen incorrect checksum while the latter, posing as a special case, zeroes the field instead, \ie this could appear as if the field is unused.

\afblock{TCP Options.}
TCP specifies up to 40 bytes of options for future extensibility.
It is thus crucial that these bytes are actually usable and, if used, handled correctly.
According to the specification, any implementation is required to support the \ac{eool}, \ac{noop}, and \ac{mss} option.
We test these options due to their significance for interoperability and, in the general case, extensibility and performance.
The different, and sometimes variable, option length makes header parsing somewhat computationally expensive (especially in hardware), opening the door for non-conformant performance enhancements comparable to skipping checksum verification. 
Further, an erroneous implementation of either requirement can have security repercussions in the form of buffer overflows or resource wastage, culminating in a denial of service. 
The \optionSupport test validates the support of \ac{eool} and \ac{noop}, while the \optionUnknown test checks the handling of an unassigned option.
The \mssSupport test verifies the proper handling of an explicitly stated \ac{mss} value, while the \mssMissing test tests the usage of default values specified by the RFC in the absence of the \ac{mss} option.

\afblock{TCP Flags.}
Alongside the stated TCP options, TCP's extensibility is mainly guaranteed by (im-)mutable control flags in its header, of which four are currently still reserved for future use.
The most prominent ``recent'' example is \ac{ecn}~\cite{floyd:rfc3168:TheAdditionofExplicitCongestionNotificationECNtoIP}, which uses two previously reserved bits. 
Though not explicitly stated as a numbered formal MUST\footnote{RFC\,793bis-Draft14 states: \textit{``Must be zero in generated segments and must be ignored in received segments, if corresponding future features are unimplemented by the sending or receiving host.''}~\cite{eddy:rfc793Draft14:tcpTransmissionControlProtocolSpecification}}, a TCP must zero (when sending) and ignore (when receiving) unknown header flags, which we test with the \reserved test, as incorrect handling can considerably block or delay the adoption of future features. 

The \urgentPointer test addresses the long-established URG flag.
Validating the support of segments flagged as urgent, the test splits around 500 bytes of urgent data into a sequence of three segments with comparable sizes. Each segment is flagged as urgent, and the urgent pointer field caries the offset from its current sequence number to the sequence number following the urgent data, \ie to the sequence number following the payload.
Initially intended to speed up segment processing by indicating data which should be processed immediately, the widely-used \ac{bsd} socket interface instead opted to interpret the urgent data as out-of-band data, leading to diverging implementations.
As a result, the urgent pointer's usage is discouraged for new applications~\cite{eddy:rfc793Draft14:tcpTransmissionControlProtocolSpecification}. 
Nevertheless, TCP implementations are still required to support it with data of arbitrary length. 
As the requirement's inclusion adds computational complexity, implementers may see an incentive to skip it. 

\afblock{Pass and Failure Condition Notation.}
For the remainder of this paper, we use the following notation to report passing or failing of the above-described tests.
Connections that unmistakably conform are denoted as \emph{PASS}, whereas not clearly determinable results (applies only to some tests) are conservatively stated as \emph{UNK}.
UNKs may have several reasons such as, \eg{}hosts ceasing to respond to non-test packets after having responded to a liveness test.
Non-conformities raised by the target host are denoted as \emph{\ftarget}, and non-conformities raised by middleboxes on the path rather than the probed host are denoted as \emph{\fpath}.

\subsection{Validation}
\label{sec:evaluation}

To evaluate our test design, we performed controlled measurements using a testbed setup, thereby eliminating possible on-path middlebox interference.
Thus, only \ftarget{} can occur in this validation, but not \fpath.
To cover a broad range of hosts, we verified our test implementations by targeting current versions of the three dominant \acp{oss} (Linux, Windows, and macOS) as well as three alternative TCP stacks (uIP~\cite{uip}, lwIP~\cite{lwip}, and Seastar~\cite{seastar}).

\begin{table}[t]
	\small
	\rowcolors{2}{gray!15}{white}
	\centering
	\begin{tabularx}{\textwidth}{l|*{6}{Y}}
	\rowcolor{white!100}
	 \multicolumn{1}{c|}{MUST Test as}						& Linux		& Windows 	& macOS 	& uIP 		& lwIP 		& Seastar \\
	 \multicolumn{1}{c|}{defined in Table~\ref{tab:requirements}}	& 5.2.10  	& 1809 		& 10.14.6 	& 1.0 		& 2.1.2		& 19.06 \\\Xhline{0.75pt}
	\checksumIncorrect 	& \cmark 	& \cmark 	& \cmark 	& \cmark 	& \cmark 	& \xmark \\
	\checksumZero 	 	& \cmark 	& \cmark 	& \cmark 	& \cmark 	& \cmark 	& \xmark \\
	\optionSupport 	 	& \cmark 	& \cmark 	& \cmark 	& \cmark 	& \cmark 	& \cmark \\
	\optionUnknown 	 	& \cmark 	& \cmark 	& \cmark 	& \cmark 	& \cmark 	& \cmark \\
	\mssSupport 	 	& \cmark 	& \xmark 	& \cmark 	& \cmark 	& \cmark 	& \cmark \\
	\mssMissing 	 	& \cmark 	& \cmark 	& \xmark 	& \cmark 	& \cmark 	& \cmark \\
	\reserved 		& \cmark 	& \cmark 	& \cmark 	& \cmark 	& \cmark 	& \cmark \\
	\urgentPointer 	 	& \cmark 	& \cmark 	& \cmark 	& \xmark 	& \cmark 	& \cmark \\
	\end{tabularx}
	\vspace{2.5pt}
	\caption{Results of testbed measurements stating PASS (\cmark) and \ftarget (\xmark)}
	\label{tab:evaluation}
	\vspace{-2em}
\end{table}

We summarize the results in Table~\ref{tab:evaluation}.
As expected, we observe a considerable degree of conformance.
Linux, as well as lwIP, managed to achieve full conformance to the tested requirements.
Surprisingly, all other stacks failed in at least one test each.
That is, most stacks do not fully adhere to these minimum requirements.
uIP exposed the most critical flaw by crashing when receiving a segment with urgent data, caused by a segmentation fault while attempting to read beyond the segment's size (see Section \ref{sec:design}).
Since the release of the tested Version of uIP, the project did not undergo further development, but instead moved to the Contiki OS project~\cite{contiki}, where it is currently maintained in Contiki-NG~\cite{contiki.ng}.
Following up on Contiki, it was uncovered that both distributions are still vulnerable.
Their intended deployment platform, embedded microcontrollers, often lack the memory access controls present in modern \acp{oss}, amplifying the risk that this flaw poses.
Addressing this issue, we submitted a Pull request to Contiki-NG~\cite{contiki.ng.pr}.
The remaining \ftarget{} have much less severe repercussions.
Seastar, which bypasses the Linux L4 network stack using \emph{Virtio}~\cite{virtio}, fails both checksum tests.
While hardware offloading is enabled by default, Seastar features software checksumming, which should take over if offloading is disabled or unsupported by the host OS.
However, host OS support of offloaded features is not verified, which can lead to mismatches between believed to be and actually enabled features.
We reported this issue to the authors \cite{seastar:issue}.
The tests pass if the unsupported hardware offloads are manually deselected.
The \ftarget{} failure for macOS in the \mssMissing test is a consequence of macOS defaulting to a 1024 bytes \ac{mss} regardless of the IP version, thereby exceeding the IPv4 TCP default \ac{mss}, and falling behind that of IPv6.
Windows 10 applies the \ac{mss} defaults defined in the TCP specification as a lower bound to any incoming value, overwriting the 515 bytes advertised in the \mssSupport test. 
Both \ac{mss} non-conformities could be mitigated by path \ac{mtu} discovery, dynamically adjusting the segment size to the real network path.

\takeaway{Only two tested stacks (Linux and lwIP) pass all tests and show full conformance. Surprisingly, all other stacks failed in at least one category each. That is, non-conformance to basic mandatory TCP implementation requirements already exists in current OS implementations.
Even though our testbed validation is limited in the OS diversity, we can already expect to find a certain level of host non-conformance when probing TCP implementations in the wild.
}

\section{TCP Conformance in the Wild}
\label{sec:results}

 In the following, we move on from our controlled testbed evaluation and present our measurement study in the Internet.
 Before we present and discuss the obtained results, we briefly focus on our measurement setup and our selected target sets.
 
\subsection{Measurement Setup \& Target Hosts}
\label{sec:setup}

\afblock{Measurement Setup.}
Our approach involves performing active probes against target hosts in the Internet to obtain a representative picture of TCP conformance in the wild.
All measurements were performed using a single vantage point within the IPv4 research network of our university between August 13 and 22, 2019.
As we currently do not have IPv6-capable scan infrastructure at our disposal, we leave this investigation open for future work.
Using a probing rate of 10k pps on a distinct 10GBit/s uplink, we decided to omit explicit loss detection and retransmission handling due to the increased complexity, instead stating results possibly affected by loss as UNK if not clearly determinable otherwise.

\afblock{Target Hosts.}
To investigate a diverse set of end-to-end paths as well as end hosts, a total of 3,731,566 targets have been aggregated from three sources: \emph{i)} the HTTP Archive \cite{online:httpArchive}, \emph {ii)} Alexa Internet's top one million most visited sites list \cite{online:alexa,toplists}, and \emph{iii)} Censys~\cite{durumeric:ccs15:censys} port 80 and 443 scans.

The \textbf{HTTP Archive} regularly crawls about 5M domains obtained from the Chrome User Experience Report to study Web performance and publishes the resulting dataset. 
We use the dataset of July 2019. 
For this, we were especially interested in the \ac{cdn} tagged URLs, as no other source provides URL-to-\ac{cdn} mappings.
Since no IP addresses are provided, we resolved the 876,835 URLs to IPv4 addresses through four different public DNS services of Cloudflare, Google, DNS.WATCH, and Cisco's OpenDNS.
Some domains contain multiple \ac{cdn} tags in the original dataset.
For these cases, we obtained the \ac{cdn} mapping from the chain of CNAME resource records in the DNS responses and excluded targets that could still not be linked to only a single \ac{cdn}.
Removing duplicates on a per-URL basis, one target per resolved IPv4 address was selected. 
The resulting 4,116,937 targets were sampled to at most 10,000 entries per \ac{cdn}, leading to 147,318 hosts in total. 
Removing duplicate IP addresses and blacklist filtering, we derived the final set of 27,795 \ac{cdn} targets.

As recent research has shown~\cite{alashwali:ares19:wwwTLS}, prefixing \texttt{www.}\ to a domain might not only provide different TLS security configurations and certificates than their non-\texttt{www} counterparts, but might also (re-)direct the request to servers of different \acp{cp}. 
To study this implications on TCP conformance, we used the \textbf{Alexa 1M list} published on August 10th, 2019, and resolved every domain with and without \texttt{www}-prefix according to the process outlined in the HTTP Archive. 
The resulting 3,297,849 targets were further sampled, randomly selecting one target with and without \texttt{www}-prefix per domain, removing duplicate IP addresses and blacklist filtering, leading to 466,685 Alexa targets.

\textbf{Censys} provided us research access to their data of Internet-wide port scans, which represent a heterogeneous set of globally distributed targets. 
In addition to the IPv4 address and successfully scanned port, many targets include information on host, vendor, OS, and product. 
Using the dataset compiled on August 8th, 2019, 10,559,985 Censys targets were identified with reachable ports 80 or 443, including, but not limited to, IoT devices, customer-premises equipment, industrial control systems, remote-control interfaces, and network infrastructure appliances. 
By removing duplicate IP addresses and blacklist filtering we arrive at 3,237,086 Censys target hosts.

\afblock{Ethical Considerations.}
We aim to minimize the impact of our active scans as much as possible.
First, we follow standard approaches~\cite{durumeric:usenix13:EthicsInternetScanning} to display the intent of our scans in rDNS records of our scan IPs and on a website with an opt-out mechanism reachable via each scan IP.
Moreover, we honor opt-out requests to our previous measurements and exclude these hosts.
We further evaluated the potential implications of the uIP/Contiki crash observed in Section~\ref{sec:evaluation}.
Embedded microcontrollers, commonly used in IoT devices, are the primary use-case of uIP/Contiki.
We could not identify hosts using this stack in the Censys device type data to exclude IPs, but assume little to very little use of this software stack within our datasets.
We thus believe the potential implications to be minimal.
We confirm this by observing that 100\% of failed targets in the CDN as well as the Alexa dataset, and 99.35\% of failed targets in the Censys dataset, are still reachable following \urgentPointer test case execution.
We thus argue that our scans have created no harm to the Internet at large.

\subsection{Results and Discussion}
\label{sec:discussion}

We next discuss the results of our conformance testing, which we summarize in Table~\ref{tab:results:overview}.
The table shows the relative results per test case for all reachable target hosts, excluding the unreachable ones.
As the target data was derived from the respective sources multiple days before executing the tests (see Section~\ref{sec:setup}), unreachable targets are expected.
Except for minor variations, which can be explained by dynamic IP address assignment and changes to host configurations during test execution, \ca12\% of targets could not be reached in each test case and are removed from the results.
While the CDN and Alexa datasets were derived from sources featuring popular websites, we expect a large overlap of target hosts, which is confirmed by 15,387 targets present in both datasets.
Alexa and Censys share only 246 target hosts, while CDN and Censys do not overlap.
All datasets are publicly available \cite{datasets}.
The decision to classify a condition as PASS, UNK, \ftarget, or \fpath, does vary between test cases as a result of their architecture (see Section~\ref{sec:design}) and are discussed in detail next.

\afblock{TCP Checksum.}
We start with the results of our checksum tests that validate correct checksum handling.
As Table~\ref{tab:results:overview} shows, CDNs have a low failure rate for both tests, and we do not find any evidence for on-path modifications.
In contrast, hosts from the Alexa and the Censys dataset show over \ca3\% \ftarget{} failures.
Drilling down on these hosts, they naturally cluster into two classes when looking at the AS ownership.
On the one hand, we find AS (\eg Amazon), where roughly 7\% of all hosts fail both tests.
Given the low share, these hosts could be purpose build high-performance VMs, \eg for TCP-terminating proxies that do not handle checksums correctly.
On the other hand, we find hosts (\eg hosted in the QRATOR filtering AS) where nearly all hosts in that AS fail the tests.
Since QRATOR offers a DDoS protection service, it is a likely candidate for operating a special purpose stack.

\takeaway{We find cases of hosts that do not correctly handle checksums.
While incorrect checksums may be a niche problem in the wild, these findings highlight that attackers with access to the unencrypted payload, but without access to the headers, could alter segments and have the modified data accepted.}

\begin{table}[t]
	\small
	 \setlength{\aboverulesep}{0pt}
   	 \setlength{\belowrulesep}{0pt}
    	\setlength\extrarowheight{0pt}
    	\setlength{\tabcolsep}{0.5pt}
	\rowcolors{2}{gray!15}{white}
	\centering
	\begin{tabularx}{\textwidth}{l|YYY|YYY|YYY}
	\rowcolor{white!100}
& \multicolumn{3}{c|}{\textbf{CDN}} 			& \multicolumn{3}{c|}{\textbf{Alexa}}		& \multicolumn{3}{c}{\textbf{Censys}}	 	\\
	\multicolumn{1}{c|}{MUST Test as}					& \multicolumn{3}{c|}{$n$ = 27,795} 				& \multicolumn{3}{c|}{$n$ = 466,685}		 	& \multicolumn{3}{c}{$n$ = 3,237,086}			\\
	\rowcolor{white!100}
	\multicolumn{1}{c|}{defined in Table~\ref{tab:requirements}}	& UNK      	& \ftarget			& \fpath 	 	& UNK       & \ftarget			& \fpath 	& UNK     	& \ftarget			& \fpath 	\\ \Xhline{0.75pt}\checksumIncorrect 	& 0.234 	& 0.374 			& -  			& 0.441 	& \textbf{3.224} 	& 0.002  	& 3.743 	&  \textbf{3.594} 	& 0.003  	\\
	\checksumZero 	 	& 0.253 	& 0.377 			& -  			& 0.455 	& \textbf{3.210} 	& 0.001  	& 3.873 	&  \textbf{3.592} 	& 0.003 	\\
	\optionSupport 	  	& - 		& 0.040 			& -  	 		& - 		& 0.470 			& 0.009  	& - 		& 1.410 			& 0.313 	\\
	\optionUnknown 	  	& - 		& 0.026 			& 0.011  		& - 		& 0.585 			& 0.053  	& - 		& 1.477 			& 0.019 	\\
	\mssSupport 	 	& -  		& 0.018 			& - 			& - 		& 0.728 			& 0.002  	& - 		& 0.412 			& 0.004 	\\
	\mssMissing 	 	& 0.026 	& -  				& 0.018  		& 0.303 	& 0.299 			& 0.136  	& 1.423 	& 0.388 			& 0.416 	\\
	\reserved 	  	& - 	    & \textbf{2.194} 	& 0.011  		& - 		& \textbf{6.689} 	& 0.293  	& - 		&  \textbf{2.791} 	& 0.048 	\\
	\textit{Reserved-SYN} 	& - 	    & 0.138 	& 0.011  		& - 		& 1.297 	& 0.309  	& - 		&  1.849 	& 0.049 	\\
	\urgentPointer 	  	& 0.150 	& 0.330 			& 0.022  		& 0.804 	& \textbf{3.179} 	& 0.208  	& 3.815 	&  \textbf{7.300} 	& 0.042 	\\ \end{tabularx}
	\vspace{2.5pt}
	\caption{Overview of relative results (in \%) per test case per dataset.
	Here, $n$ denotes the number of targets in each dataset. 
	For better readability, we do not show the PASS results and highlight excessive failure rates in bold.}
	\label{tab:results:overview}
	\vspace{-2em}
\end{table}

\afblock{TCP Options.}
We next study if future TCP extensibility is honored by the ability to use TCP options.
In our four option tests (see Table~\ref{tab:results:overview} for an overview), we observe overall the lowest failure rates---a generally good sign for extensibility support.
Again, the Censys dataset shows the most failures, and especially the \optionSupport and the \mssMissing test have the highest \fpath{} (middlebox failures) across all tests.
Both tests show a large overlap in the affected hosts and have likely the same cause for the high path failure rates.
We observe that these hosts are all located in ISP networks.
For the \mssMissing failures, we observe that an \ac{mss} is inserted at these hosts---likely due to the ISPs performing \ac{mss} clamping, \eg due to PPPoE encapsulation by access routers.
These routers need to rewrite the options header (to include the \ac{mss} option), and as the \optionSupport fails when, \eg some of the \ac{eool} and \ac{noop} are stripped, the exact number of \ac{eool} and \ac{noop} are likely not preserved.
Still, inserting the \ac{mss} option alters the originally intended behavior of the sender, \ie having an \ac{mss} of 536 byte for IPv4. In this special case, the clamping did actually increase the \ac{mss}, and thereby strip some of the \ac{eool} and \ac{noop} options.

Looking at the \optionUnknown test, where we send an option with an unallocated codepoint, we again see low \fpath{} failures, but still, a non-negligible number of \ftarget{} fails.
There is no single AS that stands out in terms of the share of hosts that fail this test. However, we observe that among the ASes with the highest failure rates are ISPs and companies operating Cable networks.

Lastly, the \mssSupport test validating the correct handling of \ac{mss} values shows comparably high conformance. 
As we were unable to clearly pinpoint the failures to specific ASes, the most likely cause can be traced to the non-conformant operating systems as shown by our validation (see Section~\ref{sec:evaluation}), where Windows fails this test and likely others that we did not test in isolation.

\takeaway{Our TCP options tests show the highest level of conformance of all tests, a good sign for extensibility. Still, we find cases of middlebox inference, mostly \ac{mss} injectors and option padding removers---primarily in ISP networks hinting at home gateways. Neither is inherently harmful due to path \ac{mtu} discovery and the voluntary nature of option padding.}

\afblock{TCP Flags.}
Besides the previously tested options, TCP’s extensibility is mainly guaranteed by (im-)mutable control flags in its header to toggle certain protocol behavior.
In the \reserved test, we identify the correct handling of \emph{unknown} (future) flags by sending an unallocated flag and expect no change in behavior.
Instead, we surprisingly observe high failure rates across all datasets, most notable CDNs.
When inspecting the CDN dataset, we found \ca10\% of Akamai's hosts to show this behavior.
We contacted Akamai, but they validated that their servers do \emph{not} touch this bit.
Further analysis revealed that the reserved flag on the SYN was truthfully ignored, but \emph{our test} failed as the final ACK of the 3-way handshake (second part of the test, see Table~\ref{tab:requirements}), which also contains the reserved flag, was seemingly dropped as we got SYN/ACK retransmissions.
However, this behavior originates from the usage of Linux's \textit{TCP\_DEFER\_ACCEPT} socket option, which causes a socket to only wakeup the user space process if there is data to be processed \cite{online:linuxTcpManPage}.
The socket will wait for the first data segment for a specified time, re-transmitting the SYN/ACK when the timer expires in the hope of stimulating a retransmission of possibly lost data.
Since we were not sending any data, we eventually received a SYN/ACK retransmission, seemingly due to the dropped handshake-completing ACK with the reserved flag set. 
Hence, we credited the retransmission to the existence of the reserved flag at first, later uncovering that the retransmission was unrelated to the reserved flag, but actually expected behavior using the \textit{TCP\_DEFER\_ACCEPT} socket option.
Following up with Akamai, they were able to validate our assumption by revealing that parts of their services utilize this socket option.
While it is certainly debatable if deliberately ignoring the received ACK is a violation of the TCP specification, our test fails to account for this corner case. 
Thus, connectivity is \textit{not} impaired.

In contrast, connectivity \textit{is} impaired in the cases where our reserved flag SYN fails to trigger a response at all, leaving the host unreachable (see \textit{Reserved-SYN} in Table~\ref{tab:results:overview}).
The difference between both failure rates thus likely denotes hosts using the defer accept mechanism, as CDNs, in general, seem to comply with the standard.
We also observe a significant drop in failures in the Alexa targets.
While our results are unable to show if \emph{only} defer accepts are the reason for this drop, they likely contribute significantly as TCP implementations would need to differentiate between a reserved flag on a SYN and on an ACK, which we believe is less likely.
Our results motivate a more focused investigation of the use of socket options and the resulting protocol configurations and behavioral changes.

Lastly, the URG flag is part of TCP since the beginning to indicate data segments to be processed immediately.
With the \urgentPointer test we check if segments that are flagged as urgent are correctly received and acknowledged.
To confirm our assumption of this test having minimal implications on hosts due to the uIP/Contiki crash (see Section \ref{sec:evaluation}), we checked if the \ftarget{} instances were still reachable after test execution.
Our results show that of these failed targets, 99.35\% of Censys, and 100\% of CDN and Alexa, did respond to our following connection requests, which were part of the subsequent test case executed several hours later.
While we argue that these unresponsive hosts can be explained by dynamic IP address assignment due to the fluctuating nature of targets in the Censys dataset, we recognize that the implicit check within the subsequent test case is problematic due to the time period between the tests and the possibility of devices and services being (automatically) restarted after crashing.
We thus posit, that future research should include explicit connectivity checks directly following test case execution on a per target basis, and skip subsequent tests if a target's connectivity is impaired.

Surprisingly, the \urgentPointer test shows the highest failure rate among all tests.
That is, segments flagged as urgent are \emph{not correctly} processed.
In other words, flagging data as urgent limits connectivity.
We find over \ca7\% of hosts failing in the Censys dataset, where ISPs again dominate the ranking.
Only about 1.2\% of these failures actively terminated the connection with a RST, while the vast majority silently discarded the data without acknowledging it.
Looking at Alexa and CDNs, we again find an Amazon AS at the top.
Here, we randomly sampled the failed hosts to investigate the kind of services offered by them.
At the top of the list, we discovered services that were proxied by a \textit{Vegur}~\cite{online:vegur}, respective \textit{Cowboy}~\cite{online:cowboy}, proxy server that seem to be used in tandem with the \textit{Heroku}~\cite{online:heroku} cloud platform.
Even though we were unable to find how Heroku precisely operates, we suspect a high-performance implementation that might simply not implement the urgent mechanism at all.

\takeaway{While unknown flags are often correctly handled, they can reduce reachability, especially when set on SYNs. The use of the urgent pointer resulted in the highest observed failure rate by hosts that do not process data segments flagged as urgent. 
Thus, using the reserved flags or setting the urgent pointer limits connectivity in the Internet.}

We therefore posit to remove the mandatory implementation requirement of the urgent pointer from the RFC to reflect its deprecation status, and thus explicitly state that its usage can break connectivity.
Future protocol standards should therefore be accompanied by detailed socket interface specifications, \eg as has been done for IPv6 \cite{stevens:rfc3542:APIforIPv6,gilligan:rfc3493:SocketInterfaceExtensionsforIPv6}, to avoid RFC misconceptions.
Moreover, we started a discussion within the IETF, addressing the issue encountered with the missing formal MUST requirement of unknown flags, which potentially led and/or will lead to diverging implementations \cite{tcpm:reserved:entry}.
Additionally, we proposed a new MUST requirement, removing ambiguities in the context of future recommended, but not required, TCP extensions which allocate reserved bits \cite{tcpm:reserved:proposal}.

\textbf{Alexa: Does www.\ matter?}
It is known that \texttt{www.domain.tld} and \newline \texttt{domain.tld} can map to different hosts~\cite{alashwali:ares19:wwwTLS}, \eg the CDN host vs.\ the origin server, where it is often implicitly assumed that both addresses exhibit the same behavior.
However, 4.89\% (11.4k) of the Alexa domains with and without \texttt{www.}\ prefix show different conformance levels to at least one test.
That is, while the host with the \texttt{www.}\ prefix can be conformant, the non-prefixed host could not, and vice versa.
Most of these non-conformance issues are caused by TCP flags, for which we have seen that they can impact the reachability of the host.
That is, 53.3\% of these domains failed the reserved flags test, and 58\% the urgent pointer test (domains can be in both sets).
Thus, a website can be unreachable using one version and reachable by the other.

\takeaway{While the majority of Alexa domains are conformant, the ability to reach a website can differ whether or not the \texttt{www.}\ prefix is used.}

\section{Conclusion}
\label{sec:conclusion}

This paper presents a broad assessment of TCP conformance to mandatory MUST requirements.
We uncover a non-negligible set of Internet hosts and paths that do not adhere to even basic requirements.
Non-conformance already exists at the OS-level, which we uncover in controlled testbed evaluations: only two tested stacks (Linux and lwIP) pass all tests.
Surprisingly, others (including macOS and Windows) fail in at least one category each.
A certain level of non-conformance is therefore expected in the Internet and highlighted by our active scans.
First, we observe hosts that do not correctly handle checksums.
Second, while TCP options show the highest level of conformance, we still find cases of middlebox inference, mostly \ac{mss} injectors and option padding removers---primarily in ISP networks hinting at home gateways.
Moreover, and most worrisome, using reserved flags or setting the urgent pointer can render the target host unreachable.
Last, we observe that 4.8\% of Alexa-listed domains show different conformance levels when the \texttt{www.}\ prefix is used, or not, of which more than 50\% can be attributed to TCP flag issues---which can prevent connectivity.
Our results highlight that conformance to even fundamental protocol requirements should not be taken for granted but instead checked regularly.

\pagebreak
\bgroup
\section*{Acknowledgments}
\label{sec:acks}
This work has been funded by the DFG as part of the CRC 1053 MAKI within subproject B1.
We would like to thank Akamai Technologies for feedback on our measurements, Censys for contributing active scan data, and our shepherd Robert Beverly and the anonymous reviewers.

\bibliographystyle{splncs04}

\begin{thebibliography}{10}
\providecommand{\url}[1]{\texttt{#1}}
\providecommand{\urlprefix}{URL }
\providecommand{\doi}[1]{https://doi.org/#1}

\bibitem{contiki.ng.pr}
{Contiki-NG TCP URG Pull Request}.
  \url{https://github.com/contiki-ng/contiki-ng/pull/1173}

\bibitem{contiki.ng}
{Contiki-NG: The OS for Next Generation IoT Devices}.
  \url{https://github.com/contiki-ng}

\bibitem{contiki}
{Contiki OS}. \url{https://github.com/contiki-os}

\bibitem{online:cowboy}
Cowboyku, \url{https://github.com/heroku/cowboyku}

\bibitem{datasets}
{Dataset to "MUST, SHOULD, DON'T CARE: TCP Conformance in the Wild"}.
  \doi{10.18154/RWTH-2020-00809}

\bibitem{online:heroku}
Heroku platform, \url{https://www.heroku.com/}

\bibitem{lwip}
{lwIP - A Lightweight TCP/IP stack}.
  \url{http://savannah.nongnu.org/projects/lwip/}

\bibitem{seastar}
{Seastar}. \url{https://github.com/scylladb/seastar}

\bibitem{seastar:issue}
{Seastar: Virtio device reports features not supported by the OS}.
  \url{https://github.com/scylladb/seastar/issues/719}

\bibitem{online:linuxTcpManPage}
tcp(7) - linux man page, \url{https://linux.die.net/man/7/tcp}

\bibitem{tcpm:reserved:entry}
{TCPM Mailinglist: RFC793bis draft 14 Reserved Bits: Problem statement}.
  \url{https://mailarchive.ietf.org/arch/msg/tcpm/s0LtY3Ce3QBBAkJ_DuSH5VDNFMY}

\bibitem{tcpm:reserved:proposal}
{TCPM Mailinglist: RFC793bis draft 14 Reserved Bits: Proposal}.
  \url{https://mailarchive.ietf.org/arch/msg/tcpm/_jpUQx0AjByR3UOgyX88RWoTxL0}

\bibitem{uip}
{uIP}. \url{https://github.com/adamdunkels/uip}

\bibitem{online:vegur}
Vegur: Http proxy library, \url{https://github.com/heroku/vegur}

\bibitem{virtio}
{Virtio: Paravirtualized drivers for kvm/Linux}.
  \url{https://www.linux-kvm.org/page/Virtio}

\bibitem{alashwali:ares19:wwwTLS}
Alashwali, E.S., Szalachowski, P., Martin, A.: {Does ``www.''\ Mean Better
  {Transport Layer Security}?} In: ACM International Conference on
  Availability, Reliability and Security (ARES) (2019).
  \doi{10.1145/3339252.3339277}

\bibitem{online:alexa}
{Alexa Internet}: About us, \url{https://www.alexa.com/about}

\bibitem{ECNReadiness}
Bauer, S., Beverly, R., Berger, A.: Measuring the state of {ECN} readiness in
  servers, clients, and routers. In: ACM Internet Measurement Conference (IMC)
  (2011). \doi{10.1145/2068816.2068833}

\bibitem{RobPassiveTCPFingerprinting}
Beverly, R.: {A Robust Classifier for Passive TCP/IP Fingerprinting}. In:
  Passive and Active Measurement Conference (PAM) (2004).
  \doi{10.1007/978-3-540-24668-8\_16}

\bibitem{bradner:rfc2119:Keywords}
Bradner, S.O.: {Key words for use in RFCs to Indicate Requirement Levels}. RFC
  2119 (Mar 1997). \doi{10.17487/RFC2119}

\bibitem{cardwell:usenix13:packetdrill}
Cardwell, N., Cheng, Y., Brakmo, L., Mathis, M., Raghavan, B., Dukkipati, N.,
  keng Jerry~Chu, H., Terzis, A., Herbert, T.: {packetdrill: Scriptable Network
  Stack Testing, from Sockets to Packets}. In: {USENIX} Anual Technical
  Conference (ATC) (2013),
  \url{https://www.usenix.org/conference/atc13/technical-sessions/presentation/cardwell}

\bibitem{rfc3234}
Carpenter, B., Brim, S.: Middleboxes: Taxonomy and issues (2002).
  \doi{10.17487/RFC3234}

\bibitem{TCPHiccups}
Craven, R., Beverly, R., Allman, M.: {A middlebox-cooperative TCP for a non
  end-to-end internet}. In: ACM SIGCOMM (2014). \doi{10.1145/2619239.2626321}

\bibitem{detal:imc13:tracebox}
Detal, G., Hesmans, B., Bonaventure, O., Vanaubel, Y., Donnet, B.: {Revealing
  Middlebox Interference with Tracebox}. In: ACM Internet Measurement
  Conference (IMC) (2013). \doi{10.1145/2504730.2504757}

\bibitem{durumeric:ccs15:censys}
Durumeric, Z., Adrian, D., Mirian, A., Bailey, M., Halderman, J.A.: {A Search
  Engine Backed by {I}nternet-Wide Scanning}. In: {ACM} Conference on Computer
  and Communications Security (CCS) (2015). \doi{10.1145/2810103.2813703}

\bibitem{durumeric:usenix13:EthicsInternetScanning}
Durumeric, Z., Wustrow, E., Halderman, J.A.: {ZMap: Fast Internet-wide Scanning
  and Its Security Applications}. In: {USENIX} Security Symposium (2013),
  \url{https://www.usenix.org/conference/usenixsecurity13/technical-sessions/paper/durumeric}

\bibitem{eddy:rfc793Draft14:tcpTransmissionControlProtocolSpecification}
Eddy, W.: {Transmission Control Protocol Specification}. Internet-Draft
  draft-ietf-tcpm-rfc793bis-14, Internet Engineering Task Force (Jul 2019),
  \url{https://datatracker.ietf.org/doc/html/draft-ietf-tcpm-rfc793bis-14},
  work in Progress

\bibitem{edeline:tma19:TransportOssification}
{Edeline}, K., {Donnet}, B.: {A Bottom-Up Investigation of the Transport-Layer
  Ossification}. In: Network Traffic Measurement and Analysis Conference (TMA)
  (2019). \doi{10.23919/TMA.2019.8784690}

\bibitem{floyd:rfc3168:TheAdditionofExplicitCongestionNotificationECNtoIP}
Floyd, S., Ramakrishnan, D.K.K., Black, D.L.: {The Addition of Explicit
  Congestion Notification (ECN) to IP}. RFC 3168 (Sep 2001).
  \doi{10.17487/RFC3168}

\bibitem{nmapFingerprinting}
Fyodor: Remote os detection via tcp/ip stack fingerprinting.
  \url{https://nmap.org/nmap-fingerprinting-article.txt} (1998)

\bibitem{gilligan:rfc3493:SocketInterfaceExtensionsforIPv6}
Gilligan, R.E., McCann, J., Bound, J., Thomson, S.: {Basic Socket Interface
  Extensions for IPv6}. RFC 3493 (Mar 2003). \doi{10.17487/RFC3493}

\bibitem{honda:imc11:ExtendTcp}
Honda, M., Nishida, Y., Raiciu, C., Greenhalgh, A., Handley, M., Tokuda, H.:
  {Is It Still Possible to Extend TCP?} In: ACM Internet Measurement Conference
  (IMC) (2011). \doi{10.1145/2068816.2068834}

\bibitem{online:httpArchive}
{HTTP Archive}: About {HTTP Archive}, \url{https://httparchive.org/about}

\bibitem{middleboxNegotiation}
Knutsen, A., Ramaiah, A., Ramasamy, A.: Tcp option for transparent middlebox
  negotiation. \url{https://tools.ietf.org/html/draft-ananth-middisc-tcpopt-02}
  (2013)

\bibitem{kuehlewind:tma18:tracingInternetPathTransparency}
{K{\"u}hlewind}, M., {Walter}, M., {Learmonth}, I.R., {Trammell}, B.: {Tracing
  Internet Path Transparency}. In: Network Traffic Measurement and Analysis
  Conference (TMA) (2018). \doi{10.23919/TMA.2018.8506532}

\bibitem{MirjaECN}
Kühlewind, M., Neuner, S., Trammell, B.: On the state of {ECN} and {TCP}
  options on the internet. In: Passive and Active Measurement Conference (PAM)
  (2013). \doi{10.1007/978-3-642-36516-4\_14}

\bibitem{LangleyECN}
Langley, A.: Probing the viability of {TCP} extensions.
  \url{http://www.imperialviolet.org/binary/ecntest.pdf} (2008)

\bibitem{BobECNPlus}
{Mandalari}, A.M., {Lutu}, A., {Briscoe}, B., {Bagnulo}, M., {Alay}, O.:
  {Measuring ECN++: Good News for ++, Bad News for ECN over Mobile}. IEEE
  Communications Magazine  \textbf{56}(3),  180--186 (March 2018).
  \doi{10.1109/MCOM.2018.1700739}

\bibitem{tfoInitialDeployment}
Mandalari, A.M., Bagnulo, M., Lutu, A.: {TCP Fast Open}: initial measurements.
  In: ACM CoNEXT Student Workshop (2015)

\bibitem{marinos:sigcomm14:NetworkStackPerformanceSpecialization}
Marinos, I., Watson, R.N., Handley, M.: {Network Stack Specialization for
  Performance}. In: ACM SIGCOMM (2014). \doi{10.1145/2619239.2626311}

\bibitem{marinos:sigcomm17:DiskCryptNet}
Marinos, I., Watson, R.N., Handley, M., Stewart, R.R.: {Disk, Crypt, Net:
  Rethinking the Stack for High-performance Video Streaming}. In: ACM SIGCOMM
  (2017). \doi{10.1145/3098822.3098844}

\bibitem{MedinaIMC2004}
Medina, A., Allman, M., Floyd, S.: {Measuring Interactions between Transport
  Protocols and Middleboxes}. In: ACM Internet Measurement Conference (IMC)
  (2004). \doi{10.1145/1028788.1028835}

\bibitem{MedinaCCR2005}
Medina, A., Allman, M., Floyd, S.: {Measuring the Evolution of Transport
  Protocols in the Internet}. SIGCOMM Comput. Commun. Rev.  \textbf{35}(2),
  37–52 (Apr 2005)

\bibitem{tfoChristoph}
Paasch, C.: Network support for tcp fast open. Presentation at NANOG~67 (2016)

\bibitem{pahdye:sigcomm01:tbit}
Padhye, J., Floyd, S.: {On Inferring TCP Behavior}. In: ACM SIGCOMM (2001).
  \doi{10.1145/383059.383083}

\bibitem{piraux:epiq18:QuicImplementationObserving}
Piraux, M., De~Coninck, Q., Bonaventure, O.: {Observing the Evolution of QUIC
  Implementations}. In: ACM CoNEXT Workshop on the Evolution, Performance, and
  Interoperability of QUIC (EPIQ) (2018). \doi{10.1145/3284850.3284852}

\bibitem{unknown:rfc793:TransmissionControlProtocol}
Postel, J.: {Transmission Control Protocol}. RFC 793 (Sep 1981).
  \doi{10.17487/RFC0793}

\bibitem{rueth:tma18:DemystifyCdnIw}
{R{\"u}th}, J., {Hohlfeld}, O.: {Demystifying TCP Initial Window Configurations
  of Content Distribution Networks}. In: Network Traffic Measurement and
  Analysis Conference (TMA) (2018). \doi{10.23919/TMA.2018.8506549}

\bibitem{iwpaper}
R\"uth, J., Bormann, C., Hohlfeld, O.: {Large-Scale Scanning of {TCP's} Initial
  Window}. In: ACM Internet Measurement Conference (IMC) (2017).
  \doi{10.1145/3131365.3131370}

\bibitem{iwTNSM}
R\"uth, J., Kunze, I., Hohlfeld, O.: {{TCP}'s Initial Window –- Deployment in
  the Wild and its Impact on Performance}. IEEE Transactions on Network and
  Service Management (TNSM)  (2019). \doi{10.1109/TNSM.2019.2896335}

\bibitem{rueth:pam19:hiddentreasures}
R{\"u}th, J., Zimmermann, T., Hohlfeld, O.: {Hidden Treasures --- Recycling
  Large-Scale Internet Measurements to Study the Internet's Control Plane}. In:
  Passive and Active Measurement Conference (PAM) (2019).
  \doi{10.1007/978-3-030-15986-3\_4}

\bibitem{toplists}
Scheitle, Q., Hohlfeld, O., Gamba, J., Jelten, J., Zimmermann, T., Strowes,
  S.D., Vallina-Rodriguez, N.: {A Long Way to the Top: Significance, Structure,
  and Stability of Internet Top Lists}. In: ACM Internet Measurement Conference
  (IMC) (2018). \doi{10.1145/3278532.3278574}

\bibitem{defeatingFingerprinting}
Smart, M., Malan, G.R., Jahanian, F.: {Defeating TCP/IP Stack Fingerprinting}.
  In: USENIX Security Symposium (2000)

\bibitem{stevens:rfc3542:APIforIPv6}
Stevens, W.R., Thomas, M., Nordmark, E., Jinmei, T.: {Advanced Sockets
  Application Program Interface (API) for IPv6}. RFC 3542 (Jun 2003).
  \doi{10.17487/RFC3542}

\bibitem{stone:sigcomm00:checksums}
Stone, J., Partridge, C.: {When the CRC and TCP Checksum Disagree}. In: ACM
  SIGCOMM (2000). \doi{10.1145/347059.347561}

\end{thebibliography}
\balance

\end{document}